\documentclass[a4paper,11pt]{article}
\usepackage{aaskaiid}
\usepackage{orcidlink}

\usepackage{enumitem}
\usepackage{float}
\usepackage{amsfonts,amsmath,mathtools,bm}
\usepackage{hyperref}
\usepackage[nameinlink,capitalise]{cleveref}
\usepackage[none]{hyphenat}
\newcommand{\HI}{\textsc{Hi}}
\newcommand{\de}{\mathrm{d}}

\title{Cosmology with Tully-Fisher \HI{} Galaxy Surveys}
\ShortTitle{Cosmology with Tully-Fisher \HI{} Galaxy Surveys}

\author[1]{Jo\"{e}l Mayor \orcidlink{0009-0008-6499-8013}}
\ShortName{J. Mayor et al.}
\emailAdd{jmayor@ethz.ch}
\author[2]{Anthony Carr \orcidlink{0000-0003-4074-5659}}
\author[3]{Khaled Said \orcidlink{0000-0002-1809-6325}}
\author[4,5]{Gabriella De Lucia \orcidlink{0000-0002-6220-9104}}
\author[6]{Anastasia Ponomareva \orcidlink{0000-0003-4100-0173}}
\author[7,8]{Philip Bull \orcidlink{0000-0001-5668-3101}}
\author[9,10,11,8]{Stefano Camera \orcidlink{0000-0003-3399-3574}}
\author[7]{Ainulnabilah Nasirudin \orcidlink{0000-0003-2213-4547}}
\author[12,8]{Marta Spinelli \orcidlink{0000-0003-0148-3254}}
\author[6]{Robert Yates \orcidlink{0000-0001-9320-4958}}

\affiliation[1]{Institute for Particle- and Astrophysics, ETH Z\"urich, Wolfgang-Pauli-Strasse 27, 8093 Z\"urich, Switzerland}
\affiliation[2]{Korea Astronomy and Space Science Institute, 776 Daedeok-daero, Daejeon 34055, South Korea}
\affiliation[3]{School of Mathematics and Physics, University of Queensland, Brisbane,
QLD 4072, Australia}
\affiliation[4]{INAF, Osservatorio Astronomico di Trieste, Via Tiepolo 11, 34131 Trieste, Italy}
\affiliation[5]{IFPU -- Institute for Fundamental Physics of the Universe, Via Beirut, 2, 34151 Trieste, Italy}
\affiliation[6]{Centre for Astrophysics Research, University of Hertfordshire, Hatfield, AL10 9AB, UK}
\affiliation[7]{Jodrell Bank Centre for Astrophysics, University of Manchester, Manchester, M13 9PL, United Kingdom}
\affiliation[8]{Department of Physics and Astronomy, University of Western Cape, Cape Town 7535, South Africa}
\affiliation[9]{Dipartimento di Fisica, Universit\`a degli Studi di Torino, Via P.\ Giuria 1, 10125 Torino, Italy}
\affiliation[10]{INFN, Sezione di Torino, Via P.\ Giuria 1, 10125 Torino, Italy}
\affiliation[11]{INAF, Osservatorio Astrofisico di Torino, Via Osservatorio 20, 10025 Pino Torinese, Italy}
\affiliation[12]{Observatoire de la C\^ote d’Azur, Laboratoire J-L Lagrange, Boulevard de l'Observatoire, Nice, France}

\abstract{The SKA Observatory will enable measurements of the Tully-Fisher relation for statistical samples of \HI{} selected galaxies out to unprecedented depths and redshifts thanks to its unique combined spatial and spectral sensitivity. This chapter explores the transformative potential of such surveys for cosmology, in particular in the field of peculiar velocity measurements. We briefly review the present observational landscape for Tully-Fisher \HI{} galaxy surveys and existing peculiar velocity datasets, and compare them with predictions for SKAO Tully-Fisher \HI{} galaxy surveys with AA* and AA4 configurations of the SKA-Mid array. We discuss the extended range of cosmology science cases covered and enabled by such surveys.}

\begin{document}
\maketitle

\section{Introduction}

    The large-scale structure of the Universe encodes fundamental information about the cosmological parameters, the nature of gravity, and the physics of dark energy.
    Traditional probes of cosmic structure, such as galaxy redshift surveys, cosmic microwave background (CMB) anisotropies, and weak gravitational lensing, have collectively established the $\Lambda$CDM concordance model with remarkable precision \citep[e.g.][]{desi_2025,planck_2020,des_2024}.
    These methods primarily trace the distribution and clustering of matter through its gravitational influence on photon trajectories or galaxy positions.
    A complementary window is offered by peculiar velocities, which refer to the deviations in the motion of galaxies or other cosmic structures from the idealized uniform expansion described by the Hubble flow.
    These velocities arise from local gravitational interactions, where matter over-densities induce accelerations that perturb objects away from purely radial recession.
    As a result, peculiar velocities provide observational probes of the cosmic velocity field, measuring the time derivative of the matter distribution rather than the matter distribution itself.
    In cosmology, this distinction provides unique leverage for constraining the growth rate of structure as a function of the cosmic scale factor, $f(a) \equiv \mathrm{d}\ln D/\mathrm{d}\ln a$, that is sensitive to modifications of General Relativity and to the properties of dark energy \citep[e.g.][]{guzzo_2008,huterer_2017}.
    Moreover, peculiar velocities are unbiased tracers of the underlying total matter density, enabling cleaner tests of gravitational physics than clustering measurements alone.
    Despite this promise, peculiar velocity cosmology has historically been limited by small sample sizes ($N \lesssim 10^4$ galaxies), shallow redshift coverage ($z \lesssim 0.1$), and heterogeneous datasets combining multiple distance indicators with different systematic uncertainties \citep[e.g.][]{springob_2014,tully_2016,howlett_2022}.
    This has limited the cosmological impact of velocity surveys relative to their density-based counterparts, leaving much of the velocity field's potential untapped.

    The Tully-Fisher relation \citep{tully_fisher_1977}, a tight empirical correlation between galaxy luminosity (or baryonic mass) and rotation velocity, serves as the crucial bridge connecting disk galaxy kinematics to cosmological distance measurements.
    Initially developed as a redshift-independent distance indicator, the Tully-Fisher relation has evolved into a prominent method for deriving peculiar velocities across local volumes.
    Its baryonic formulation \citep[e.g.][]{mcgaugh_2000,bell_2001,mcgaugh_2005,lelli_2016b,lelli_2019}, relating total baryonic mass to a characteristic rotation velocity, exhibits remarkably small intrinsic scatter when calibrated with well-resolved rotation curves, making it a precision tool for distance estimation \citep{verheijen_2001,ponomareva_2017,lelli_2016a}.
    The 21\,cm emission line from neutral hydrogen (\HI{}) plays a central role in this application: its interferometric observations simultaneously yield the systemic redshift, the integrated line flux (tracing \HI{} mass), and the velocity width (probing rotational kinematics).
    When combined with optical or near-infrared photometry for stellar mass and inclination measurements, \HI{} observations provide all ingredients necessary for Tully-Fisher distance estimation and subsequent peculiar velocity derivation.
    Beyond its utility as a local distance indicator, the Tully-Fisher relation offers direct tests of galaxy formation physics and dark matter--baryon coupling across cosmic time \citep{dutton_2012,dave_2019,baes_2025}.
    Recent observations extending to $z \sim 0.35$ suggest that the baryonic Tully-Fisher relation remains remarkably stable over the past four billion years \citep{jarvis_2025}, providing confidence that it can be reliably calibrated and applied at intermediate redshifts.
    This stability, combined with the dual role of \HI{} as kinematic tracer and redshift indicator, positions Tully-Fisher \HI{} surveys as an optimal pathway for large-scale peculiar velocity measurements.

    The SKA Observatory (SKAO) will have the potential to unlock a new frontier in peculiar velocity cosmology by providing an unprecedented combination of sensitivity, angular resolution, and survey volume for \HI{} observations.
    Compared to existing pathfinder facilities such as the Australian SKA Pathfinder (ASKAP) and the Meer-Karoo Array Telescope (MeerKAT), SKA-Mid will allow for homogeneous \HI{} detections for up to an order of magnitude more galaxies, extending up to $z \sim 0.4$.
    This extended redshift reach is cosmologically significant: it encompasses roughly 4 billion years of cosmic history, spanning the transition between cosmic epochs of matter domination to dark energy domination.
    The combination of large sky coverage (e.g.~$5,000$ deg$^2$), and high angular and spectral resolution sufficient to resolve double-peaked \HI{} profiles will yield well-characterized rotation curves for individual galaxies, enabling robust inclination corrections and velocity width measurements.
    Crucially, the SKAO era coincides with a suite of complementary optical and near-infrared surveys covering the southern hemisphere:
    the \textsl{Vera C. Rubin Observatory's Legacy Survey of Space and Time} (LSST; \citealt{ivezic_2019});
    the \textsl{Euclid} mission \citep{laureijs_2011,euclid_2025};
    and spectroscopic programs such as the \textsl{4-metre Multi-Object Spectroscopic Telescope Hemisphere Survey} (4HS; \citealt{taylor_2023}).
    These will provide deep multiwavelength photometry, accurate photometric and spectroscopic redshifts, stellar mass estimates, and morphological measurements for essentially all SKAO \HI{} detections.
    This synergy removes a historical bottleneck in Tully-Fisher cosmology:  heterogeneous, incomplete ancillary optical data.
    With simultaneous access to \HI{} kinematics and optical properties for statistically complete samples, SKAO Tully-Fisher surveys will enable precision measurements of the expansion rate $H_0$, of the growth rate $f\!\sigma_8(z)$, of velocity/momentum field clustering statistics, of bulk flows on scales exceeding 100 Mpc, and of a plethora of cross-correlations with other cosmic probes, delivering a comprehensive picture of structure formation in the low-to-intermediate redshift Universe.

    This chapter explores the potential of Tully-Fisher \HI{} galaxy surveys with the SKAO for cosmology.
    We structure it with, first (\autoref{sec:landscape}), a brief overview of the current state of the field, to give a qualitative context.
    Second (\autoref{sec:forecasts}), we produce forecasts of SKAO capabilities for a Tully-Fisher \HI{} galaxy survey, which provide a complementary quantitative context.
    Third (\autoref{sec:science_cases}), we consider and discuss an extensive suite of cosmological science cases enabled by large-scale Tully-Fisher surveys with the SKAO.\\
    In more detail, we review the physical basis of the Tully-Fisher relation (\autoref{sec:TF}), the landscape of existing and ongoing \HI{} surveys (\autoref{sec:surveys}), the state-of-the-art methodology for deriving peculiar velocities from Tully-Fisher distances (\autoref{sec:TF_PVs}), and the current state of peculiar velocity measurements (\autoref{sec:PVcat}).
    We obtain instrument sensitivity limits for the SKAO (\autoref{sec:sensitivity}), from which we derive projected number counts (\autoref{sec:counts}), and discuss unique advantages and synergies with optical surveys (\autoref{sec:synergies}).
    Finally, \autoref{sec:discussion} summarizes the challenges, limitations, systematic uncertainties and mitigation strategies.

\section{\texorpdfstring{Tully-Fisher \HI{} Galaxy Surveys}{Tully-Fisher HI Galaxy Surveys}} \label{sec:landscape}


\subsection{The Tull-Fisher Relation}\label{sec:TF}

    The Tully-Fisher relation (TFr; \citealt{tully_fisher_1977}) is one of the most fundamental empirical laws in extragalactic astronomy. It links the luminosity of a late-type galaxy  to its rotational velocity, establishing a tight correlation between its visible baryonic content and its total dynamical mass, dominated by dark matter.

    In essence, the TFr maps the virial relationship between rotational kinetic energy and gravitational binding energy in dark-matter-dominated, rotationally-supported systems. The flat outer rotation curve, traced by extended \HI{} discs, reflects the asymptotic circular velocity of the host dark matter halo, fixed in turn by the virial scaling between halo mass and circular velocity \citep{mo_1998}. The existence of a tight correlation between this velocity and the baryonic content of the disc is then a consequence of the remarkable structural regularity of late-type galaxies, with discs assembling at near-universal central baryonic surface densities \citep{freeman_1970}: haloes of similar circular velocity end up hosting baryonic discs of similar mass. That this correlation persists across the diverse merger and accretion histories of late-type galaxies points to the action of self-regulating baryonic processes, which collectively erase much of the variance one would naively expect in the baryon-to-halo connection. The TFr therefore admits a natural interpretation as the projection onto observables of the dynamical equilibrium between dark matter haloes and the rotationally supported baryonic discs they host.
    
    In observational terms, luminosity, particularly in the near-infrared, traces the stellar mass that represents the bulk of the baryonic content for massive systems \citep{ponomareva_2018}, while the rotational velocity reflects the depth of the total gravitational potential, i.e. the dark matter mass. At low rotational velocities (i.e. for gas-dominated dwarfs), the stellar component is a minor fraction of the baryonic mass. Including the contribution from cold gas leads to the baryonic Tully-Fisher relation (bTFr; \citealt{mcgaugh_2000, lelli_2016b, lelli_2019}), which relates the total baryonic mass ($M_\star + M_{\rm gas}$) to the characteristic rotational velocity, typically measured in terms of the 21\,cm spectral line width of \HI{}. This form of the relation extends over nearly five orders of magnitude in mass and exhibits remarkably small intrinsic scatter when based on well-resolved \HI{} rotation curves \citep{verheijen_2001, ponomareva_2017, lelli_2016a}. The slope and tightness of the bTFr provide stringent tests for cosmological simulations, challenging models of baryon–dark-matter coupling, feedback efficiency, and angular-momentum retention \citep{dutton_2012, glowacki_2020, baes_2025}.
    
    The TFr and its baryonic form offer a unique bridge between galaxy evolution and cosmology. If the relation evolves with redshift, its zero point and slope encode changes in the efficiency of baryonic structure assembly within dark matter haloes. Measuring any such evolution directly tests cosmological simulations of galaxy growth \citep{dave_2019} and informs models of star-formation feedback and gas accretion. Conversely, a lack of evolution would imply that the physical processes establishing the baryon–halo connection were already in place by $z \approx 1$. Studies using molecular (CO) or optical (H$\alpha$) tracers have yielded mixed results \citep{topal_2018, tiley_2019}, but these tracers probe only the inner disc regions and do not fully capture the gravitational potential traced by extended \HI{} discs \citep{frank_2016}. Hence, robust conclusions about the evolution of the bTFr require spatially resolved \HI{} observations over a cosmologically significant redshift range.
    
    Current limitations arise primarily from sensitivity and resolution. \HI{} emission becomes increasingly faint with redshift, making detections beyond $z \approx 0.1$ challenging for existing interferometers. This restricts direct bTFr studies to the nearby Universe and forces reliance on heterogeneous data or alternative tracers beyond the local Universe. Furthermore, accurate velocity deprojection requires reliable inclination measurements from optical or near-infrared imaging, adding an additional source of uncertainty. In addition, observational biases, such as preferential detection of massive, gas-rich systems, can distort the inferred slope and scatter, complicating comparisons with simulations. Consequently, a homogeneous and statistically robust \HI{} sample is essential for the exploitation of the TFr for cosmology.


    While the 21\,cm line of \HI{} has emerged as a key tracer for Tully-Fisher measurements, optical spectroscopy has a long history in this field and continues to play an important role.
    Early Tully-Fisher studies relied on optical rotation curves obtained from long-slit spectroscopy of H$_\alpha$ or other emission lines \citep{rubin_1985,mathewson_1992,vogt_1997}.

    The advent of multiplexed spectroscopic surveys has revitalized interest in optical Tully-Fisher measurements.
    Most notably, the \textsl{Dark Energy Spectroscopic Instrument} (DESI) is undertaking an ambitious program to measure Tully-Fisher distances \citep{saulder_2023,douglass_2025}.
    DESI's approach involves placing multiple fibers across the major axis of each target galaxy, measuring rotation velocities from the redshift difference between fiber positions.
    This technique represents a significant advance in efficiency over traditional long-slit spectroscopy, enabling wide-area surveys that complement northern hemisphere coverage where \HI{} surveys have limited reach.

    However, optical spectroscopy Tully-Fisher measurements face several inherent challenges with respect to the \HI{}-based approaches.
    First, optical rotation curves probe only the inner disk regions where stellar light dominates, missing the flat portion of the rotation curve that extends into the dark matter-dominated outer disk traced by \HI{} \citep{frank_2016}.
    This limitation introduces additional scatter and potential systematic biases in the Tully-Fisher relation.
    Second, the fiber-based approach requires galaxies to be sufficiently large and favorably oriented to accommodate multiple fiber placements.
    This requirement becomes increasingly restrictive at higher redshifts where angular sizes diminish.
    Third, the measured rotation velocity depends sensitively on the galactocentric radius of fiber placement, introducing an additional source of systematic uncertainty.
    Fourth, optical measurements require careful correction for stellar absorption, emission line contamination, and dust extinction effects that vary with wavelength and galactic position.
    
    In contrast, \HI{} observations provide several decisive advantages.
    The 21\,cm line traces the full extent of the gaseous disk, capturing the flat tail of the rotation curve that provides the most robust dynamical mass estimate.
    A single \HI{} spectrum yields both the systemic velocity and the full velocity width, eliminating uncertainties associated with spatial sampling in optical surveys.
    Radio observations are unaffected by dust extinction, a significant advantage for edge-on systems where optical measurements are most compromised despite inclinations corrections being minimal.
    Moreover, \HI{} measurements enable homogeneous, untargeted surveys that detect all \HI{}-rich galaxies above the flux limit within the survey volume.
    This eliminates the complex selection biases inherent to targeted optical surveys, where pre-selection based on photometric properties can introduce subtle systematic effects that propagate through to cosmological inferences.
    Untargeted surveys thus carry a striking efficiency advantage over e.g. DESI, which requires careful targeting and multiple fiber placements per galaxy, with associated overhead for acquisition and configuration.
    Each \HI{} detection automatically provides all necessary kinematic information without the need for follow-up observations.
    Furthermore, \HI{} interferometric observations enable spatially resolved \HI{} maps for a substantial fraction of detections, providing both kinematic and morphological inclinations from the same dataset.
    These fundamental advantages position \HI{} observations as the primary path toward large Tully-Fisher surveys.
    
\subsection{\texorpdfstring{\HI{} Galaxy Surveys}{HI Galaxy Surveys}}\label{sec:surveys}

    The past two decades have seen remarkable progress in untargeted \HI{} galaxy surveys, transitioning from pioneering single-dish observations to interferometric campaigns with SKAO pathfinder instruments, such as the Australian SKA Pathfinder (ASKAP) and the Meer-Karoo Array Telescope (MeerKAT).
    These surveys provide the observational foundation for Tully-Fisher cosmology, extending previous work to fainter flux densities, larger volumes, and higher redshifts.
    Here we review the key past and ongoing surveys that have shaped the current state-of-the-art and established the methodology for future SKAO observations.
    \autoref{tab:HIsurvey_comparison} and \autoref{tab:HIdeepsurvey_comparison} present a condensed overview of their respective specifications.

\subsubsection{Wide-field surveys}

    \begin{table*}
    \centering
    \caption{Landscape of wide-field untargeted \HI{} galaxy surveys:\\ (a) \citet{meyer_2004}, (b) \citet{haynes_2018}, (c) \citet{zhang_2024}, (d) \citet{koribalski_2020}.}
    \label{tab:HIsurvey_comparison}
    \begin{tabular}{lrrcccc}
        \hline
        Survey & Sample & Area & Channel sensitivity & Channel width & Max.~$z$ & Ref. \\
         &  & [deg$^2$] & [mJy beam$^{-1}$] & [km s$^{-1}$] &  & \\
        \hline
        HIPASS & $4'315$ & $21'341$ & $\sim13$ & $\sim18$ & 0.04 & (a) \\
        ALFALFA & $31'502$ & $\sim7'000$ & $\sim2.0$ & $\sim10$ & 0.06 & (b) \\
        FASHI & $\sim100'000$ & $\sim22'000$ & $\sim1.5$ & $\sim6.4$ & 0.09 & (c) \\
        WALLABY & $\sim500'000$ & $\sim30'940$ & $\sim1.6$ & $\sim4$ & 0.26 & (d) \\
        \hline
        
    \end{tabular}
    \end{table*}

    \textsl{HIPASS:} The \HI{} Parkes All Sky Survey \citep{barnes_2001} was conducted with the 64-m Parkes radio telescope and provided \HI{} detections for 4,315 galaxies across the southern hemisphere over 21,341 deg$^2$ \citep{meyer_2004}.
    Operating at 1.4 GHz with a 13-beam multibeam receiver, HIPASS achieved an RMS sensitivity of $\sim$13 mJy beam$^{-1}$ per 18 km\,s$^{-1}$ channel, detecting galaxies primarily within the local Universe ($z \lesssim 0.04$).
    Despite its modest sensitivity by modern standards, HIPASS established the \HI{} mass function in the nearby volume and provided crucial calibration samples for early Tully-Fisher studies.
    The survey demonstrated the power of blind \HI{} searches and set the stage for more ambitious multibeam surveys.

    \textsl{ALFALFA:} The Arecibo Legacy Fast ALFA survey \citep{giovanelli_2005, haynes_2018} utilized the 305-m Arecibo telescope with a 7-beam L-band receiver to conduct the most extensive single-dish \HI{} survey to date.
    Operating from 2005 to 2012, ALFALFA detected $\sim$31,000 extragalactic \HI{} sources over 7,000 deg$^2$ of high Galactic latitude sky, reaching an RMS sensitivity of $\sim$2.0 mJy beam$^{-1}$ per 10 km\,s$^{-1}$ channel.
    The survey covered redshifts $z < 0.06$, with a median around $z \sim 0.02$, and provided both integrated \HI{} line fluxes and velocity widths ($W_{50}$) suitable for Tully-Fisher analysis.
    While designed primarily for \HI{} mass function studies, the large statistical sample of ALFALFA has enabled Tully-Fisher distance measurements for several thousand galaxies \citep[e.g.][]{ball_2023}, yielding constraints on local large-scale flows.
    The homogeneous nature of ALFALFA data has made it a standard reference for calibrating scaling relations and understanding selection effects in \HI{} surveys.

    \textsl{FASHI:} The Five-hundred-meter Aperture Spherical Telescope (FAST) \citep{nan_2011, li_2018}, located in Guizhou, China, is the largest single-dish radio telescope (500~m aperture) in the world at the time of writing.
    Since starting operations in 2016, FAST has conducted both targeted observations and drift-scan \HI{} surveys, including the FAST All Sky \HI{} Survey \citep[FASHI;][]{zhang_2024}, an ongoing direct successor of the ALFALFA program.
    FASHI leverages FAST's modern instrumentation and wide effective collecting area to achieve a spectral sensitivity of $\sim1.5$ mJy beam$^{-1}$ over a $\sim6.4$ km s$^{-1}$ channel width.
    
    \textsl{WALLABY:} The wide-field ASKAP L-band Legacy All-sky Blind surveY \citep{koribalski_2020} is an ongoing survey with the Australian Square Kilometre Array Pathfinder (ASKAP), representing the most ambitious wide-area \HI{} survey currently underway.
    WALLABY aims to detect $\sim$500,000 \HI{} galaxies over $\sim$75\% of the sky ($\delta < +30^\circ$; approximately 30,940 deg$^2$) out to $z \sim 0.26$, using ASKAP's 36-antenna array with a 30-deg$^2$ instantaneous field of view provided by phased array feeds.
    Operating in the 1.13--1.43 GHz band with 18.5 kHz channels ($\sim$4 km\,s$^{-1}$ velocity resolution), WALLABY achieves an RMS channel sensitivity of $\sim$1.6 mJy\,beam$^{-1}$ with $\sim$8 hours per pointing.
    The homogeneous selection function, wide redshift baseline, and large sky coverage of the survey position it as the definitive pathfinder for SKAO-scale Tully-Fisher cosmology.

\subsubsection{Deep surveys}

    \begin{table*}
    \centering
    \caption{Landscape of deepfield untargeted \HI{} galaxy surveys:\\ (a) \citet{maddox_2021}, (b) \citet{blyth_2016}, (c) \citet{duffy_2012}, (d) \citet{fernandez_2013,fernandez_2016}.}
    \label{tab:HIdeepsurvey_comparison}
    \begin{tabular}{lrcccc}
        \hline
        Survey & Area & Channel sensitivity & Channel width & Max.~$z$ & Ref. \\
         & [deg$^2$] & [$\mu$Jy beam$^{-1}$] & [km s$^{-1}$] &  & \\
        \hline
        MIGHTEE-\HI{} & $\sim32$ & $\sim100$ & $\sim5.5$ & $\sim0.58$ & (a) \\
        LADUMA & $\sim1$ & $\sim10$ & $\sim5.5$ & $\sim1.4$ & (b) \\
        DINGO (Deep) & $\sim150$ & $\sim85$ & $\sim4$ & $\sim0.26$ & (c) \\
        DINGO (Ultra-Deep)& $\sim60$ & $\sim38$ & $\sim4$ & $\sim0.43$ & (c) \\
        CHILES & $\sim0.25$ & $\sim75$ & $\sim6.6$ & $\sim0.45$ & (d) \\
        \hline
    \end{tabular}
    \end{table*}

    The new generation of deep, untargeted \HI{} galaxy surveys, such as MIGHTEE-\HI{} \citep{jarvis_2016,maddox_2021}, LADUMA \citep{blyth_2016}, DINGO \citep{meyer_2009,duffy_2012}, and CHILES \citep{fernandez_2013}, are paving the way by extending TFr/bTFr studies to progressively higher redshifts and diverse environments \citep[e.g.][]{fernandez_2016, hess_2019, ponomareva_2021, jarvis_2025}.
    These surveys provide critical insights into the evolution of the \HI{} mass function and bTFr across cosmic time, and crucial tests for galaxy formation models during an epoch when baryon assembly and feedback are most active.
    Interferometric imaging from SKAO pathfinders, ASKAP and MeerKAT, enable spatially resolved kinematics, inclination measurements from \HI{} morphology, and robust separation of rotation from random motions.
    Although their modest volumes/area restrict their application to large-scale cosmological studies, deep surveys still serve as premier testbeds for techniques and systematics at higher redshifts, relevant to Tully-Fisher cosmology with future SKAO surveys.
    
\subsection{Tully-Fisher Distances and Peculiar Velocities}\label{sec:TF_PVs}

    The TF/bTF relation yields a distance indicator, which, combined with spectroscopic redshift information enables peculiar velocity measurements.
    Hereafter we present an overview of the state-of-the-art methodology to derive such measurements, based on a wide-field untargeted \HI{} galaxy survey.
    In particular, this would be the standard approach for building a SKAO peculiar velocity survey.

    \begin{enumerate}[label=\alph*.]

    \item \HI{} Observations: Single-dish or interferometric observations yield 21\,cm spectra.
    From these spectra, via automated or semi-automated fitting the following are obtained: integrated 21\,cm flux $S_{21}$ (as $M_{\HI{}} \propto D^2 S_{21\mathrm{cm}}$); line widths $W_{20}$ or $W_{50}$ (at 20\% or 50\% of peak flux density); systemic velocity $V_{\mathrm{sys}}$ (usually taken as the midpoint of $W_{50}$).

    \item Inclination and Magnitude Measurements: Optical and/or near-infrared imaging (the latter is less affected by dust extinction) provides magnitudes in corresponding bands, and morphology information (e.g. axis ratios $b/a$) from which inclinations of individual galaxies can be derived. 

    \item Calibration: For each galaxy, the TF relation associates the line width to an absolute magnitude ($M$). This association requires a calibration using galaxies with both observed magnitudes ($m$), and independent distance measurements, including: Cepheid variables; Tip of the Red Giant Branch; Surface Brightness Fluctuations (SBF, for early-type galaxies); Type Ia supernovae.

    \item Distance \& Peculiar Velocity Estimation: Once anchored, the TF relation can be used to compute a distance modulus $\mu=M_\mathrm{TF}-m$. This, modulo any corrections related to, e.g., dust extinction, provides an estimate of the true luminosity distance $\bar{d}_{L}$. The measured distance differs from the one we compute from a cosmological model at the observed redshift ($\mu_{\mathrm{model}}(z)$) due to the galaxy's peculiar velocity (and the scatter in the TFr). Peculiar velocities can thus be inferred through a Taylor expansion of the residuals at the observed redshift, $\Delta\mu=\mu-\mu_{\mathrm{model}}(z)$, which leads to the following estimator \citep{hui_2006,carreres_2023}:
    \begin{equation}\label{eq:vp_carreres_est}
    v_\mathrm{p} = -\frac{\ln(10)c}{5}\left(\frac{(1+z)c}{H(z)d(z)}-1\right)^{-1}\Delta\mu,
    \end{equation}
    where $d(z)$ is the comoving distance calculated at $z$.
    \end{enumerate}

    Clearly, the method requires an excellent control of systematics for accurate measurements of peculiar velocities, especially in the regime where peculiar velocities are insignificant compared to recession velocities.
    This regime is reached around $z \sim 0.1$, where a typical peculiar velocity of 300 km s$^{-1}$ represents about 1\% of the recession velocity.
    Additionally, since the recovered peculiar velocities scale roughly linearly with distance (eq.~\ref{eq:vp_carreres_est}), the signal is quickly overwhelmed by TFr noise ($\Delta\mu$ scatter), though this can be mitigated with the statistical power of a large, uniform survey.
    A statistically robust sample with controlled systematics will thus enable higher redshift peculiar velocity measurements.

\subsection{Peculiar Velocity Catalogs}\label{sec:PVcat}

    Similar to the methodology outlined in \autoref{sec:TF_PVs}, other distance indicators enable peculiar velocity measurements by comparing observed redshifts with independently determined distances.
    In particular, the `fundamental plane' \citep[][]{djorgovski_1987,dressler_1987} provides a complementary distance indicator for early-type galaxies by relating their effective radii, surface brightnesses, and central velocity dispersions.
    In essence, the fundamental plane maps the virial relationship between kinetic energy, traced by dispersion, and gravitational binding energy in pressure-supported systems.
    While they target two fundamentally distinct galaxy populations, FP and TF peculiar velocity measurements represent the two main (and thus highly synergetic) options for building large catalogs.

    The current observational landscape, summarized in \autoref{tab:pvsurvey_comparison}, highlights the maturity of Tully-Fisher peculiar velocity measurements as a cosmological tool.
    However, the ongoing and near-future state-of-the-art remains fundamentally limited to the local Universe.

    \begin{table*}
    \centering
    \caption{Peculiar velocity catalogs}
    \label{tab:pvsurvey_comparison}
    \begin{tabular}{lrccr}
        \hline
        Catalog & Sample Size & Distance indicator & Redshift & Ref. \\
        \hline
        6dFGSv & $8'885$ & FP & $z<0.06$ & \cite{springob_2014}\\
        SDSS & $34'059$ & FP & $z<0.1$ & \cite{howlett_2022}\\
        CosmicFlows 4 & $55'877$ & mixed & $z<0.1$ & \cite{tully_2023}\\
        \qquad$\hookrightarrow$ & $9'984$ & BTF  & $z<0.05$ & \cite{kourkchi_2022}\\
        WALLABY & $\sim 200'000$ & TF  & $z<0.07$ & \cite{courtois_2023b}\\
        DESI & $\sim 186'000$ & FP($\sim$70\%) \& TF($\sim$30\%) & $z<0.15$ & \cite{saulder_2023}\\
        4HS & $\sim 450'000$ & FP  & $z<0.15$ & \cite{taylor_2023}\\
        \hline
        SKA-Mid AA* & $\sim 1'295'000$ & TF/BTF  & $z<0.4$ & This work: (\autoref{sec:forecasts})\\
        SKA-Mid AA4 & $\sim 2'010'000$ & TF/BTF  & $z<0.4$ & This work: (\autoref{sec:forecasts})\\
        \hline
    \end{tabular}
    \end{table*}

    The modest redshift reach of current \HI{} surveys arises from sensitivity limitations: even with ASKAP's pathfinder capabilities, \HI{} detections beyond $z \sim 0.1$ require prohibitively long integration times for wide-area coverage. This redshift barrier constrains both the accessible volume (limiting studies of large-scale structure and bulk flows to scales $\lesssim 200\,\mathrm{Mpc}$) and the cosmic time baseline (preventing direct tests of structure growth evolution).

    On the optical side, DESI TF measurements also present difficulties in observing beyond $z\sim{0.15}$ as the galaxies must be large enough to place two fibres along the axis of rotation, along with constraints of the more involved targeting regime and rotation velocity depending on the galactocentric radius at which the fibres are placed.
    
    A transformative step forward requires a facility combining order-of-magnitude improvements in sensitivity, combined with a large collecting area and wide fields of view. These are precisely the design specifications of the SKA Observatory.
    
\section{\texorpdfstring{SKAO Capabilities for Tully-Fisher \HI{} Galaxy Surveys}{SKAO Capabilities for Tully-Fisher HI Galaxy Surveys}} \label{sec:forecasts}

    To make dedicated forecasts for SKAO capabilities for Tully-Fisher and peculiar velocity measurements, we take advantage of the framework presented in \citet{mayor_2026}. The work presents a pipeline to generate realistic mock \HI{} galaxy catalogs, model a selection function, and make dedicated predictions for \HI{} galaxy surveys with the SKAO.
    
    This framework is based on outputs of semi-analytic models, building on the procedure first presented in \citet{obreschkow_2009}. Specifically, in the present work, we take advantage of the GAlaxy Evolution and Assembly (GAEA) model  \citep[][see also \citet{Ronconi01.2026.SKA} for an overview of the \HI{} simulation landscape]{de_lucia_2014,hirschmann_2016,xie_2017,fontanot_2020,xie_2020,de_lucia_2024,fontanot_2025}, and associate a modeled 21\,cm line profile to simulated galaxies.
    This allows us to make predictions for detections for the double-peaked features of the 21\,cm line, and the associated line width parameters (e.g. $W_{50}$, $W_{20}$). 
    
    All details of the theoretical framework, as well as predictions for line parameter statistics, scaling relations (e.g. the TFr), and survey selection systematics, are validated against available observational data \citep{mayor_2026}.
    
\subsection{Sensitivity limits and selection function} \label{sec:sensitivity}

\begin{figure}
        \centering
        \includegraphics[width=\linewidth]{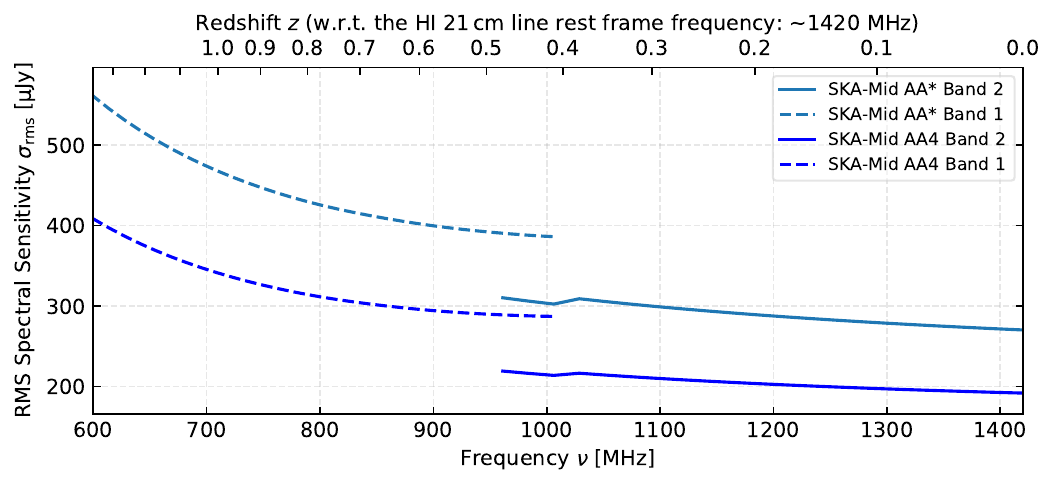}
        \caption{RMS spectral sensitivity forecasts, obtained with the SKAO sensitivity calculator API, assuming 1~hour integration time. Light blue curves and blue curves show predictions for AA*, respectively AA4 SKA-Mid configurations. Solid curves distinguish predictions for Band 2 (950 - 1760 MHz) receivers, from those for Band 1 (350 - 1050 MHz) represented with dashed curves.}
        \label{fig:rmsnoise}
    \end{figure}

    We obtain the expected sensitivity limits based on SKA-Mid AA* and AA4 receiver and array specifications\footnote{\url{https://www.skao.int/sites/default/files/documents/SKAO-TEL-0000818-V2_SKA1_Science_Performance.pdf}} using the SKAO sensitivity calculator\footnote{\url{https://www.skao.int/en/science-users/ska-tools/493/ska-sensitivity-calculators}} API.
    \autoref{fig:rmsnoise} shows the estimated RMS spectral sensitivities $\sigma_\mathrm{rms}$ for Band 1 (350\,-\,1050 MHz) and Band 2 (950\,-\,1760 MHz) receivers, assuming 1~hour integration time.

    Using these specifications, we model the signal-to-noise ratio (SNR) of the 21\,cm line width detections as \citep[e.g.][]{haynes_2018}:
    \begin{equation}\label{eq:SNR}
        \mathrm{SNR} = \frac{S_{21}}{W_{50}} \frac{\sqrt{w_\mathrm{smo}(z)}}{\sigma_\mathrm{rms}(z)}\quad ; \quad w_\mathrm{smo}(z) = \frac{W_{50}}{n\cdot |\delta V(z)|} \quad ; \quad n = 8
    \end{equation}
    where $S_{21}/W_{50}$ is an estimate of the mean flux density across the line width, and $w_\mathrm{smo}(z)$ is a ``smooth'' constraint on the number $n$ of spectral resolution bins $|\delta V(z)| = \frac{c}{\nu_e}\delta\nu (1+z)$ covering the line width, with frequency channel width $\delta\nu$.
    While surveys like ALFALFA \citep{haynes_2018} and FASHI set $n=2$, we take a more conservative approach inspired by source identification requirements for the WALLABY pre-pilot and pilot surveys \citep{courtois_2023b}, namely a detection over a minimum of 8 spectral channels.

    In the following, we use $\nu_e = 1420.40575$ MHz (rest frame \HI{} 21\,cm emission line frequency) and $\delta\nu = 13.44$ corresponding to the division of a 200 MHz band in $14'880$ channels \citep{braun_2019}.
    This yields $|\delta V(z)| \simeq 2.84 (1+z)$~km/s.
    
    We model the selection function assuming a detection limit at $\mathrm{SNR} = 5$.

\subsection{Projected number counts} \label{sec:counts}

    Following \citet{Nasirudin01.2026.SKA}, we assume the \HI{} galaxy survey design proposed in \citet{redbook_2020}, namely a Medium-Deep Band 2 SKA-MID survey covering $5'000$~deg$^2$ with a total integration time of about $t_\mathrm{tot} = 10'000$ hours. The estimated time per pointing is of $t_p = 0.95$ hours \citep[see Table 1 in][]{Nasirudin01.2026.SKA}.
    We adjust the corresponding values for $\sigma_\mathrm{rms}(z)$ using the SKAO sensitivity calculator.

    We then apply the SNR  selection cuts described in \autoref{sec:sensitivity} to a simulated galaxy catalog generated from projecting modeled galaxies and associated \HI{} lines on a mock lightcone as described in \citet{mayor_2026}.

\begin{figure}
        \centering
        \includegraphics[width=\linewidth]{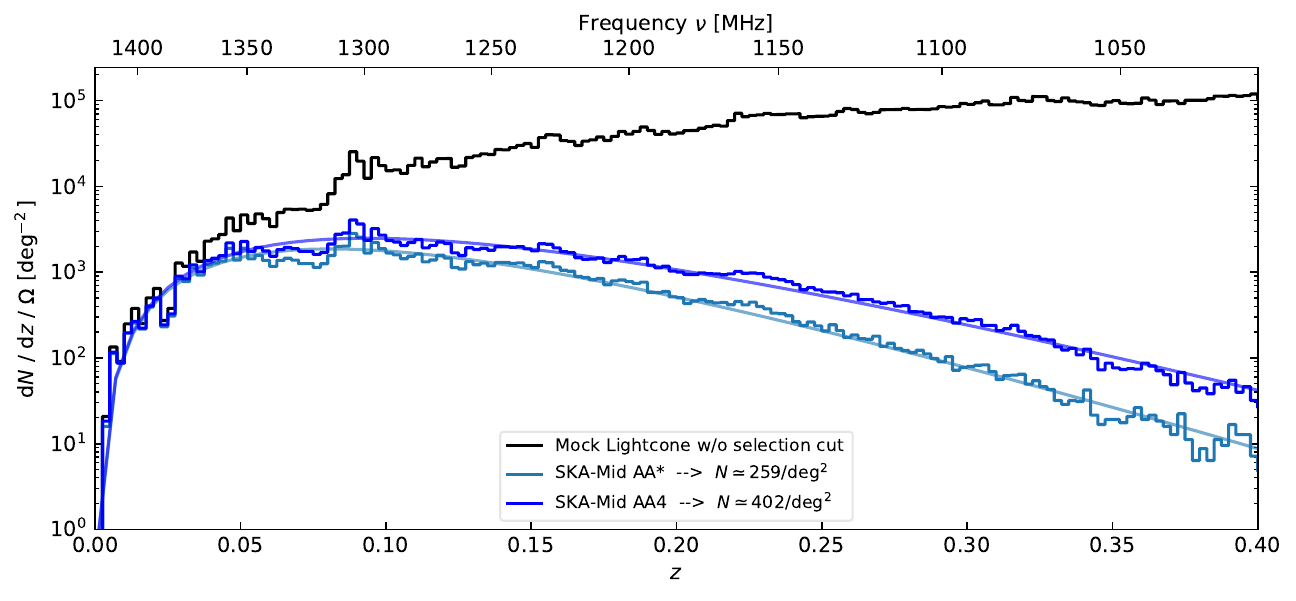}
        \caption{$\mathrm{d}N/\mathrm{d}z$ forecasts for \HI{} 21\,cm line width detections, adapted from \citet{mayor_2026}, normalized by survey area. The dark histogram shows the redshift distribution of all simulated galaxies in the mock lightcone constructed from outputs of the GAEA semi-analytic model. The light blue and blue histograms show the result obtained by applying survey selection functions described in \autoref{sec:counts}, assuming spectral sensitivities corresponding to a 10'000 hour, 5'000 deg$^2$ survey for AA*, respectively AA4 configurations of SKA-Mid. Curves of matching color show in each case a model best fit which is integrated over the redshift range to obtain estimated number counts.}
        \label{fig:dNdz}
    \end{figure}

    \autoref{fig:dNdz} shows the obtained forecast for the redshift distribution of sources $\mathrm{d}N/\mathrm{d}z$, normalized by survey area. For both SKA-Mid AA* and AA4 configurations, we fit a simple model \citep[e.g.][]{obreschkow_2009}
    \begin{equation}\label{eq:dndz}
        \frac{\mathrm{d}N/\mathrm{d}z}{\Omega} = 10^{c_1} z^{c_2} \exp(-c_3z)
    \end{equation}
    and integrate over a redshift range up to $z=0.4$ to obtain the total number of expected detections. The best fit parameters are reported in \autoref{tab:dndz_bestfit}, and the integrated number counts are listed at the bottom of  \autoref{tab:pvsurvey_comparison} for comparison with existing peculiar velocity measurements. Indeed, as TF/bTF galaxy samples are typically limited by \HI{} detections rather than by their photometric counterparts, we assume that every 21\,cm line width detection passing our selection function will yield a distance and a peculiar velocity estimate.
    While probably optimistic, this assumption is supported by unique advantages of the SKAO design and incredible synergy potential with state-of-the-art optical/NIR programs, which we discuss next.

    \begin{table*}
    \centering
    \caption{Best fit $\mathrm{d}N/\mathrm{d}z$ model parameters (\autoref{eq:dndz}) for forecasts presented in \autoref{fig:dNdz}.}
    \label{tab:dndz_bestfit}
    \begin{tabular}{lccc}
        \hline
        SKAO configuration & $c_1$ & $c_2$ & $c_3$ \\
        \hline
        SKA-Mid AA* & 6.77 & 2.30 & 27.99\\
        SKA-Mid AA4 & 6.68 & 2.25 & 23.70\\
        \hline
    \end{tabular}
    \end{table*}
    
\subsection{Unique advantages and Synergies} \label{sec:synergies}

    The interferometric array capabilities of the SKAO fundamentally distinguishes it from single-dish radio telescopes such as Arecibo or FAST.
    For \HI{}, the SKAO will obtain spectroscopic information through spectral-line imaging, which provides both spatial and kinematic information for each detected source.  These data can be used to derive simultaneous measurements of rotation velocities and inclination corrections. This reduces significantly possible systematics related to the use of external data, an important advantage given the accurate control of the systematics that is needed for peculiar velocity measurements (see \autoref{sec:TF_PVs}). 

    Optical/near-infrared photometry will still be needed to derive estimates for luminosities and/or stellar masses of the detected sources, as well as (photometric) redshift confirmation for \HI{} detections near RFI.  Fortunately,  the SKAO era will coincide (or shortly follow) with major optical/near-infrared surveys providing ideal complementary data. 
    Specifically:
    \begin{description}[font=\textsl]
    \item[LSST] \citep{ivezic_2019} will provide deep ($r \sim 27.5$) multicolor imaging over $\sim$18,000 deg$^2$ in the southern hemisphere, providing photometric redshifts, morphologies, and stellar masses for essentially all SKA \HI{} detections.
    \item[Euclid] \citep{laureijs_2011,euclid_2025} is collecting deep near-infrared imaging and slitless spectroscopy over 15,000 deg$^2$ from space, providing exquisite morphologies (this information is particularly valuable for inclination corrections), stellar masses, and accurate photometric redshifts.
    \item[4HS] \citep{taylor_2023} will collect optical spectroscopy for $\sim$30 million galaxies in the southern hemisphere, providing accurate redshifts and star formation diagnostics, along with Fundamental Plane \citep{djorgovski_1987, dressler_1987} peculiar velocity measurements for $\sim 450'000$ galaxies.
    \end{description}
    4HS in particular provides fantastic cosmological synergy.
    Most science cases of an SKAO TF survey (see \autoref{sec:science_cases}) will be enhanced when combined with the comparably sized 4HS fundamental plane sample. 

\section{Science Cases for Cosmology with an SKAO Tully-Fisher Survey} \label{sec:science_cases}

    A large, medium-deep SKAO TF \HI{} galaxy survey will enable a rich program of unique and unprecedented observational cosmology science cases, in particular through statistical analyses of the peculiar velocity field but also through cross-correlation of the cosmological information encoded in the Tully-Fisher with other probes. Hereafter, we discuss key science cases.

\subsection{\texorpdfstring{Expansion Rate $H_0$}{Expansion Rate}} \label{sec:H0}

    The expansion rate of the Universe, $H_0$, can be measured by comparing cosmological redshifts and distances. As explained above, \HI{} observations can give simultaneously a distance estimate through the TFr and an observed redshift. Usually, Type Ia supernovae are used as distance indicators to measure $H_0$ because they are relatively precise after standardisation and can be seen to great distances. However, they are relatively rare and require dedicated spectroscopic follow-up and/or time-series light-curve observations. Spiral galaxies, even only those satisfying the requirements for a successful TF measurement (e.g., not face-on, required 21 cm S/N, etc.), are abundant, providing a large statistical significance that strongly outweighs the scatter in the TFr. 
    
    As an example, $H_0$ was measured from nearly 10,000 bTFr galaxies in CosmicFlows4 with a statistical uncertainty of ${\sim}0.20$ km s$^{-1}\,\mathrm{Mpc}$$^{-1}$ \citep{kourkchi_2022}, which is a large  improvement over the ${\sim}0.95$ km s$^{-1}\,\mathrm{Mpc}$$^{-1}$ statistical uncertainty from the 277 SH0ES Type Ia supernovae \citep{riess_2022}. 
    An SKAO TF survey will deliver peculiar velocity measurements for a sample of ${\sim}10^6$ galaxies out to $z \lesssim 0.4$, potentially reducing statistical uncertainty by a factor of 10 compared to CF4. Like supernovae, the TFr only provides relative distances that need to be calibrated to an absolute scale. This is generally a major source of statistical and systematic uncertainty.
    Assuming these calibration and other systematic uncertainties, including some that might be currently unknown, can be minimized, an SKAO TF survey will easily outperform the precision of supernovae when used as a probe of local expansion history. 

\subsection{Statistics of the peculiar velocity field} \label{sec:PVstatistics}

    Peculiar velocities provide a highly complementary tracer of large-scale structure to traditional galaxy density surveys, as they arise from a fundamentally distinct cosmological perturbation: momentum transfer in the underlying matter fluid.
    While both density fluctuations and velocities trace the gravitational potential, they do so through different equations of motion.
    Poisson's equation governs density perturbations, while Euler's equation describes velocity flows.
    This distinction enables joint analyses to lift degeneracies in cosmological parameter estimation, constrain the growth rate of structure formation, and probe potential deviations from general relativity and standard cold dark matter predictions \citep[see also the related chapter in this book,][]{Camera01.2026.SKA}.

    On large scales, the evolution of cosmic structure can be described using perturbation theory.
    Treating the matter distribution as an irrotational fluid, gravitational instability is fully characterized by two fields: the density contrast, $\delta$, and the velocity divergence, $\theta\coloneqq\bm\nabla\cdot\bm v$.
    Both fields can be expanded about their linear solutions, which correspond to time-dependent scalings of the initial density field set by primordial fluctuations.
    In Fourier space, these perturbative expansions take the form
    \begin{align}
        \delta(\bm k,z)&=\sum_{m=1}^\infty D^m(z)\,\delta^{(m)}(\bm k)\;,\\
        \theta(\bm k,z)&=-\mathcal{H}(z)\,f(z)\,\sum_{m=1}^\infty D^m(z)\,\theta^{(m)}(\bm k)\;,\label{eq:expansion_theta}
    \end{align}
    where superscript $(m)$ denotes the $m$th perturbative order, $D$ is the linear growth factor normalised to unity at $z=0$, $\mathcal{H}(z) = H(z)/(1+z)$ the conformal Hubble factor, and $f(z)\coloneqq-\de\ln D/\de\ln(1+z)$ the linear growth rate.
    At first order in perturbation theory, the continuity equation imposes $\theta^{(1)}=\delta^{(1)}$, yielding the fundamental linear-theory relation $\theta=-\mathcal{H}\,f\,D\,\delta$ between velocity divergence and density contrast.

    Thanks to these statistical properties, we can construct summary statistics such as a power spectrum for peculiar velocities.
    The observable, in this case, is usually identified as the peculiar velocity projected on the line of sight, \(u\equiv v_\|\), which can be inferred from Tully-Fisher distance measurements (see \autoref{sec:TF_PVs}).
    The radial velocity power spectrum, $P_{uu}(k)$, is related to the matter power spectrum $P_{\delta\delta}(k)$ through
    \begin{equation}\label{eq:Puu}
        P_{uu}(k,z) = \left[\mathcal{H}(z)\,f(z)\right]^2 \frac{\mu^2}{k^2} P_{\delta\delta}(k,z)\;,
    \end{equation}
    where $\mu = \hat{\bm k} \cdot \hat{\bm r}$ is the cosine of the angle between the wavevector $\bm k$ and the line of sight, \(\hat{\bm r}\).
    This angular dependence introduces anisotropy in the velocity power spectrum, analogous to redshift-space distortions in galaxy clustering, and provides direct sensitivity to the growth rate parameter combination $f\!\sigma_8(z)\coloneqq f(z)\,D(z)\,\sigma_8$.

    In practice, constructing the velocity power spectrum from a Tully-Fisher survey requires careful treatment of observational systematics.
    The scatter inherent to the TFr introduces correlated noise that must be distinguished from cosmic signal.
    Furthermore, the radial velocity field samples only one component of the full three-dimensional velocity field, necessitating reconstruction techniques or forward-modeling approaches to extract cosmological constraints.
    Despite these challenges, the velocity power spectrum has been scrutinized as a powerful cosmological probe in an extensive literature over the last decade \citep{koda_2014,hellwing_2014,ivarsen_2016}.
    The volume and depth of SKAO Tully-Fisher surveys will enable unprecedented velocity power spectrum measurements, providing stringent tests of structure formation across cosmic time.


    The cosmic momentum power spectrum provides a complementary statistical probe that encodes information from both the density and velocity fields simultaneously \citep[e.g.][]{howlett_2019,qin_2019,qin_2025}.
    It can be calculated from the momentum field, $\bm\rho$, which is essentially the galaxy-weighted velocity field given by
    \begin{equation}
        \bm\rho (\bm r) = [1 + \delta_{\rm g}(\bm r)]\,\bm v(\bm r),
    \end{equation}
    where $\delta_{\rm g} (\bm r)$ and $\bm v(\bm r)$ are the galaxy overdensity and velocity at  location $\bm r$, respectively. 
    In a real survey, we are only able to measure the radial peculiar velocities of galaxies $u(\mathbf{r})$, which means that we are also only able to construct the radial momentum field p. 
    These quantities can then be assigned to a grid based on the galaxy position, and the radial momentum power spectrum at scale $\bm k$ can thus be calculated following
    \begin{equation}
    \begin{split}
        (2 \pi)^3\,\delta^D(\bm k - \bm k')\,P_p(\bm k) &= \langle[1 + \delta_{\rm g} (\bm k)]\,u(\bm k) \,[1 + \delta_{\rm g} (\bm k')]\,u(\bm k')  \rangle  \\
        &= \langle u(\bm k)\,u(\bm k') \rangle + \langle u(\bm k)\,\delta_{\rm g} (\bm k')\,u(\bm k') \rangle + \langle \delta_{\rm g} (\bm k)\,u(\bm k)\,u(\bm k') \rangle \\
        &+ \langle \delta_{\rm g} (\bm k)\,u(\bm k) \delta_{\rm g} (\bm k')\,u(\bm k') \rangle,
    \end{split}
    \end{equation}
    with $\delta^D$ being the Kronecker delta \citep{park_2000}.
    Expanding this expression reveals that the momentum power spectrum contains contributions from the velocity auto-correlation (first term), velocity-density cross-correlations (second and third terms), and a fully non-linear term coupling density and velocity at two points (fourth term).
    In linear theory, the momentum power spectrum simplifies considerably, but the presence of higher-order terms means it retains sensitivity to non-linear gravitational evolution and galaxy bias in ways distinct from either the density or velocity power spectra alone.
    
    The physical motivation for the momentum power spectrum lies in its optimal combination of density and velocity information.
    While the velocity power spectrum $P_{uu}(k)$ is free from galaxy bias at linear order, it suffers from large shot noise due to the stochastic nature of individual velocity measurements.
    Conversely, the galaxy density power spectrum $P_{\delta_{\rm g}\delta_{\rm g}}(k)$ has high signal-to-noise but is degraded by uncertain galaxy bias.
    The momentum power spectrum interpolates between these extremes: the density weighting reduces velocity shot noise, while the velocity component provides direct gravitational information. This property makes $P_p(k)$ particularly valuable for joint constraints on the growth rate $f$ and galaxy bias $b$ \citep{howlett_2019}.

    For SKAO Tully-Fisher surveys, the momentum power spectrum represents a natural analysis framework.
    The combination of \HI{} line widths (yielding peculiar velocities) and source positions (tracing the density field) from the same observations provides an internally consistent dataset for momentum field reconstruction.
    The large sample sizes anticipated will reduce shot noise to subdominant levels, allowing measurement of $P_p(\bm k)$ across the linear and mildly non-linear regimes.
    Combined with optical/NIR galaxy surveys providing higher-density tracers (e.g., from LSST, Euclid or 4HS), cross-momentum power spectra between different galaxy populations could further constrain redshift-dependent bias evolution and test assumptions about the galaxy-halo connection.
    In this context, the momentum power spectrum offers a powerful and theoretically motivated approach to extracting maximal cosmological information from SKAO peculiar velocity measurements.

\subsection{\texorpdfstring{Growth Rate $f\!\sigma_8(z)$}{Growth Rate}} \label{sec:fsigma8}

    The growth rate of cosmic structure, quantified by $f(z)$, traces the rate at which density perturbations collapse under gravity.
    This fundamental parameter is commonly expressed in its normalised form $f\!\sigma_8(z)$, where $\sigma_8$ denotes the root mean square amplitude of matter density fluctuations within spheres of radius $8\,h^{-1}\,\mathrm{Mpc}$.
    Different theories of gravity predict distinct histories for $f\!\sigma_8$, making it one of the most useful observables to distinguish between cosmological models \citep{linder_2005}.
    In General Relativity with a cosmological constant, the growth rate follows the empirical relation $f(z) \simeq \Omega_{\rm m}(z)^\gamma$, where the growth index $\gamma = 6/11 \simeq 0.55$ \citep{wang_1998, linder_2005}.
    Modified gravity theories, such as $f(R)$ gravity or the DGP model \citep{dvali_2000}, predict different values of $\gamma$, providing a direct pathway to test gravitational physics on cosmological scales \citep{linder_2007}. Any deviation from the predicted value of $\gamma = 0.55$ would provide a vital clue to modifying the theory and selecting alternatives.

    Peculiar velocity measurements offer a particularly powerful probe of $f\!\sigma_8$, and they provide several key advantages over traditional redshift-space distortion (RSD) analyses \citep{koda_2014}. First, peculiar velocities directly trace the underlying matter field, independent of galaxy bias, while RSD measurements are sensitive to the degenerate combination of bias and growth rate. Second, the velocity field is more sensitive to larger-scale modes than the density field. Third, joint analyses of density and velocity fields can significantly reduce cosmic variance.

    The combination of galaxy redshifts and peculiar velocities from distance indicators such as the Tully-Fisher relation has been demonstrated to improve constraints on $f\!\sigma_8$ by factors of 2-3 compared to RSD measurements alone in the same volume \citep{qin_2019, boubel_2024b}.
    These improvements arise from the complementary nature of the two tracers: while the density field provides high signal-to-noise on small scales, the velocity field breaks degeneracies and extends the constraining power to larger scales \citep{howlett_2017b}.
    Although SKAO Tully-Fisher \HI\ galaxy surveys will cover a smaller survey area compared to all-sky surveys, the extended redshift reach out to $z \sim 0.4$ will allow measurements of how the growth rate varies across the critical transition between matter domination and dark energy domination.
    This will provide a powerful test of whether General Relativity remains valid on scales up to billions of light-years, and offer unprecedented opportunities to constrain the growth index $\gamma$.

    \begin{figure}
    \centering
    \includegraphics[width=0.9\textwidth]{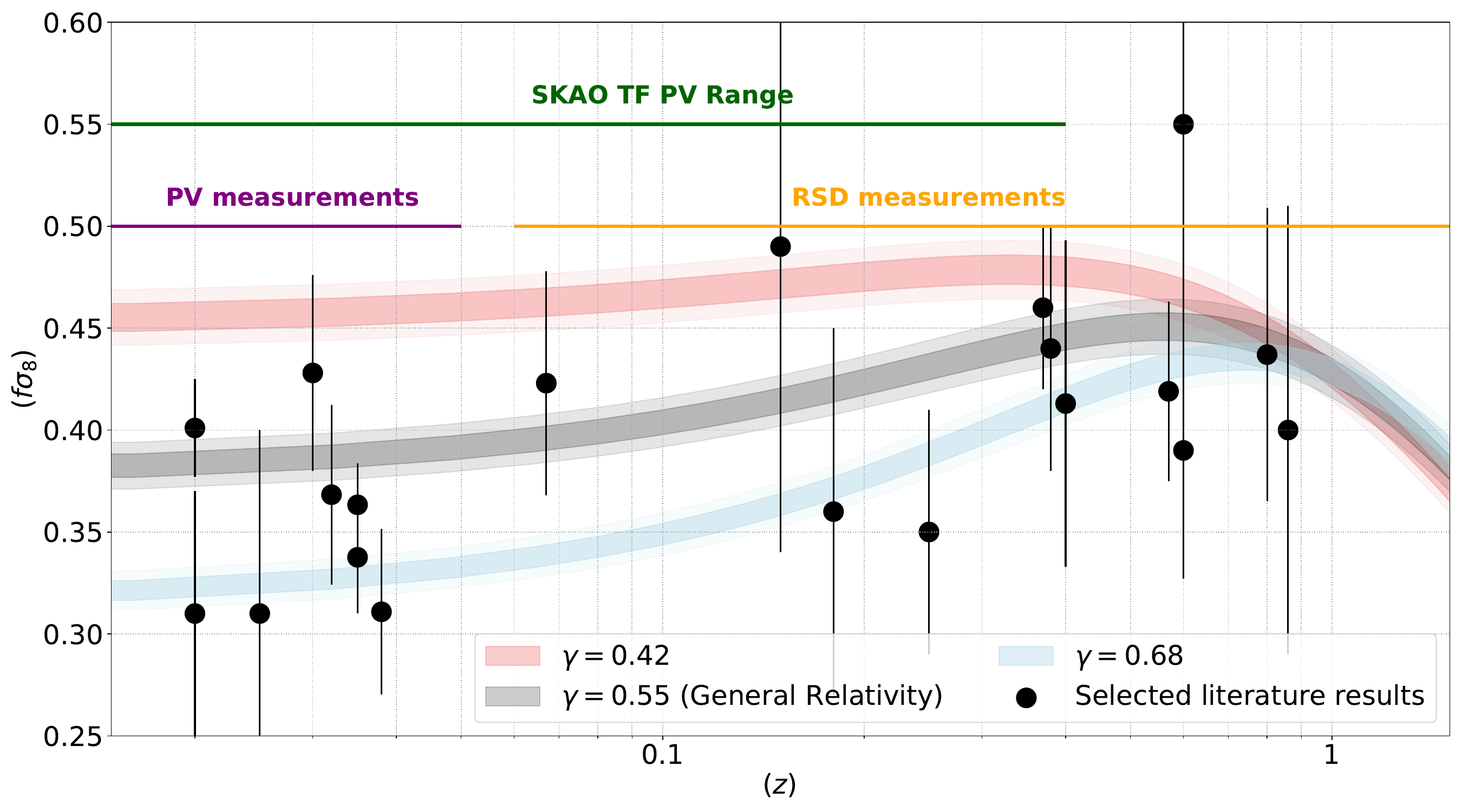}
    \caption{Growth rate of structure measurements from galaxy surveys at different effective redshifts. Black points with error bars represent selected results from the literature: measurements at $z < 0.05$ are derived from peculiar velocity surveys, while those at $z > 0.05$ come from redshift-space distortion analyses. The shaded bands show predicted growth rates for different theories of gravity using the parameterization $f(z) \simeq \Omega_{\rm m}(z)^\gamma$ \citep{linder_2007}, assuming a flat-$\Lambda$CDM cosmology. The grey band corresponds to General Relativity ($\gamma = 0.55$), while the red and blue bands show predictions for $\gamma = 0.42$ and $\gamma = 0.68$, respectively. The horizontal lines indicate the redshift ranges accessible to different observational approaches: purple for previous peculiar velocity measurements, orange for redshift-space distortions, and green for SKAO Tully-Fisher peculiar velocity surveys. The SKAO range bridges the gap between traditional low-redshift peculiar velocity studies and higher-redshift RSD measurements, enabling independent tests of gravity theories across an unprecedented redshift baseline.}
    \label{fig:testing_gr}
    \end{figure}

    \Cref{fig:testing_gr} illustrates the complementary nature of peculiar velocity and redshift-space distortion measurements in constraining the growth rate across cosmic time. Previous peculiar velocity surveys have been limited to effective redshifts $z < 0.05$, providing crucial low-redshift constraints but leaving a gap before the realm of RSD measurements, which typically dominate at $z > 0.1$. SKAO Tully-Fisher \HI{} galaxy surveys will bridge this critical gap, extending peculiar velocity measurements to $z \sim 0.4$ and thereby overlapping substantially with the RSD regime. This overlap is crucial as it will enable independent cross-validation of growth rate measurements using fundamentally different methodologies in the same redshift range. In addition, the cross-correlation of the velocity and density fields in an extended overlap region will further increase the precision with which we can measure $f\!\sigma_8$, as already demonstrated at low-$z$. The ability to measure $f\!\sigma_8$ using peculiar velocities at redshifts where RSD has traditionally been the only viable probe represents a significant advance in our ability to test General Relativity and constrain alternative gravity theories.

\subsection{Bulk Flows and Cosmological Principle} \label{sec:bulkflow}

    The bulk flow, defined as the average peculiar velocity within a spherical volume of radius $R$, provides a fundamental test of the cosmological principle, which states that the Universe is statistically homogeneous and isotropic on sufficiently large scales. In the standard $\Lambda$CDM model, peculiar velocities are generated by gravitational instability acting on density perturbations, and the bulk flow amplitude is predicted to decrease with increasing radius $R$ as we approach the scale of homogeneity.

    Recent measurements have revealed bulk flows significantly larger than these predictions, creating increasing tension with the standard cosmological model.
    Using the CosmicFlows-4 catalogue and the minimum variance estimator method, \cite{watkins_2023} reported a bulk flow of $395 \pm 29$ km s$^{-1}$ within a sphere of radius $150 \,h^{-1}\,\mathrm{Mpc}$ (equivalent to ${\sim}200\,\mathrm{Mpc}$ for $h=H_0/100\text{ km s}^{-1}\text{ Mpc}^{-1} \simeq 0.7$), with only a $\sim$0.015 per cent probability of occurring in Planck-based $\Lambda$CDM.
    Even more strikingly the measured bulk flow at $200 \,h^{-1}\,\mathrm{Mpc}$ is $427 \pm 37$ km s$^{-1}$, with a vanishingly small probability of $\sim 1.49 \times 10^{-4}$ per cent of occurring in the concordance cosmological model.
    Contrary to expectations, the bulk flow components continue to increase with radius, with no evidence of decreasing beyond $\sim 100 \,h^{-1}\,\mathrm{Mpc}$.
    This behaviour is particularly difficult to reconcile with the standard model and raises fundamental questions about either our local cosmic environment or the validity of the cosmological principle on these scales \citep{courtois_2023a, whitford_2023}.

    Despite decades of peculiar velocity surveys, the connection between these local bulk flows and the dipole observed in the cosmic microwave background has not been fully understood, nor is there consistency about the volume over which the bulk flow arises.
    SKAO surveys will be crucial to resolve these tensions \citep[see also:][]{Bertacca01.2026.SKA}.
    Extending peculiar velocity measurements out to $z \sim 0.4$ will enable bulk flow measurements over volumes approaching $\sim$1 Gpc in diameter.
    At these scales, $\Lambda$CDM predicts that the bulk flows should diminish to amplitudes well below 100 km s$^{-1}$. This will provide a definitive test of whether the observed large-scale flows converge to the cosmic microwave background rest frame or persist to scales that would fundamentally challenge our cosmological framework. Furthermore, the homogeneity and unprecedented sample size will allow detailed studies of bulk flow anisotropy and scale-dependence, potentially revealing whether these anomalously large flows arise from local structures or from more fundamental physics affecting structure growth on the largest observable scales \citep{boubel_2025a}. While the survey design ($5,000 \deg^2$) discussed in this chapter might hamper these efforts, having reduced error bars out to e.g.~$600\,\mathrm{Mpc}$ and good control over systematics could already yield important information for this field.

    \begin{figure}
    \centering
    \includegraphics[width=0.9\textwidth]{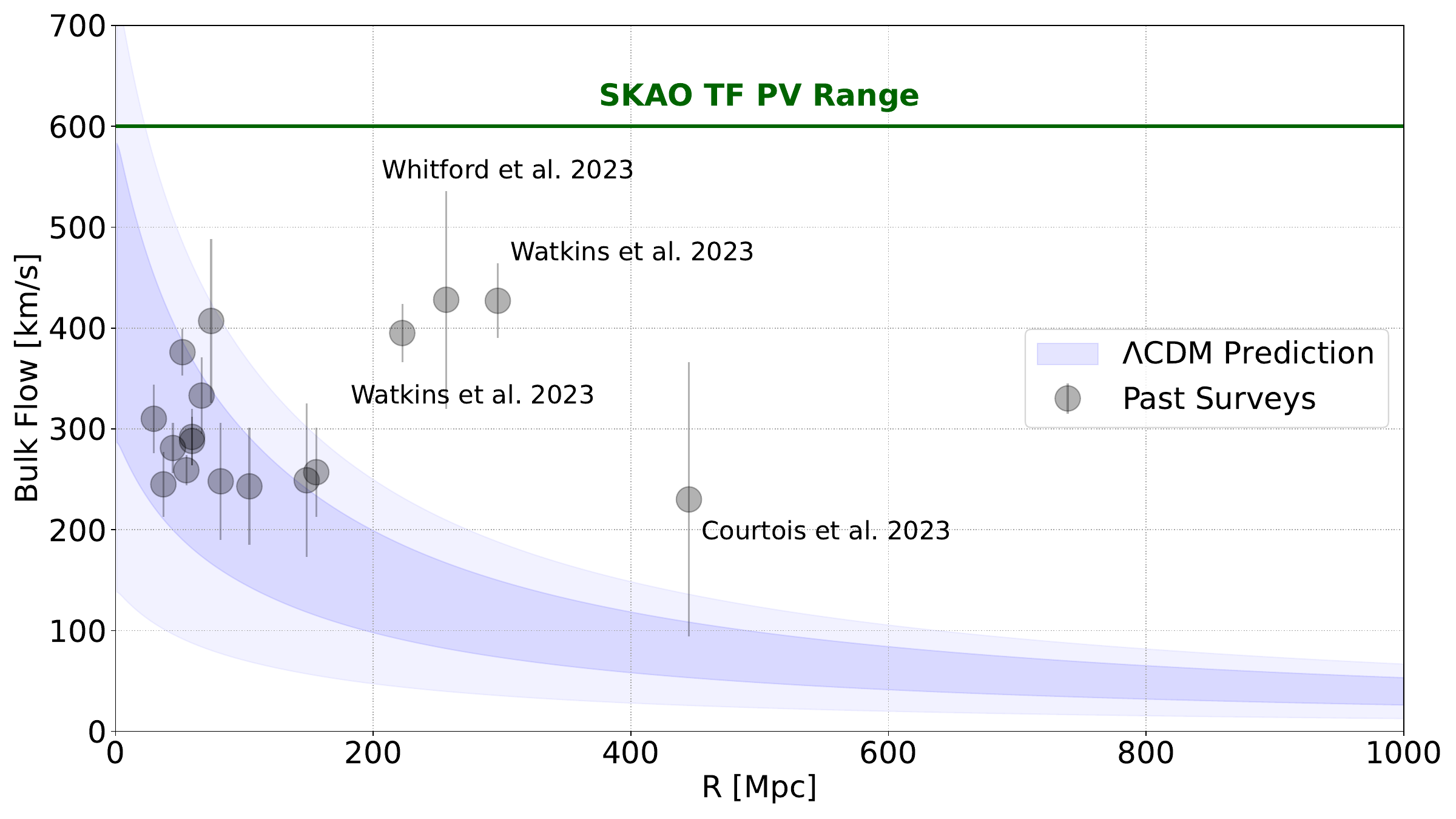}
    \caption{Bulk flow amplitude as a function of scale radius. Black points show measurements from various peculiar velocity surveys in the literature, with the three most recent analyses using CosmicFlows-4 highlighted \citep{watkins_2023, courtois_2023a, whitford_2023}. The shaded region represents the prediction from the $\Lambda$CDM model. Recent measurements show bulk flows significantly larger than predicted, with amplitudes remaining elevated beyond $200 \,h^{-1}\,\mathrm{Mpc}$ rather than decreasing as expected. The green horizontal line indicates the extended range accessible to SKAO Tully-Fisher peculiar velocity surveys, reaching scales of $\sim$1 Gpc where $\Lambda$CDM predicts bulk flows should effectively vanish, enabling a crucial test of large-scale homogeneity.}
    \label{fig:bulk_flow}
    \end{figure}

    \Cref{fig:bulk_flow} shows the evolution of bulk flow measurements as a function of scale, comparing observations from various surveys with the $\Lambda$CDM prediction. While the standard model predicts a steady decline in bulk flow amplitude with increasing radius, recent measurements using CosmicFlows-4 show persistently high values extending to $200 \,h^{-1}\,\mathrm{Mpc}$ \citep{watkins_2023, courtois_2023a, whitford_2023}. SKAO Tully-Fisher \HI{} galaxy surveys will extend bulk flow measurements to unprecedented scales approaching $\sim$1 Gpc (indicated by the green horizontal line), more than doubling the volume over which current measurements are reliable. At these scales, $\Lambda$CDM predicts bulk flows should be negligible, providing a definitive test of whether the anomalously large local flows represent a genuine cosmological tension or a statistical fluctuation in our local cosmic neighbourhood.

\subsection{3D peculiar velocity field reconstructions} \label{sec:3Dreconstruction}

    Peculiar velocity measurements from Tully-Fisher surveys provide line-of-sight velocity information at discrete galaxy positions, yielding a discrete scalar probe of the underlying 3D peculiar velocity field.
    The reconstruction of the latter is fundamentally a problem of statistical inference: Given noisy, sparse data subject to observational selection effects, what is the most likely realization of the underlying continuous velocity field consistent with both the observations and theoretical priors from cosmological structure formation?

    Several complementary approaches have been developed to address this inverse problem.
    Wiener filtering provides an optimal linear estimator that minimizes the mean-square error between the reconstructed and true fields, balancing information from data against prior expectations from the cosmological power spectrum \citep{zaroubi_1995,zaroubi_1999}.
    This technique has been successfully applied to reconstruct velocity and density fields from redshift surveys and peculiar velocity catalogs, providing smooth, volume-filling maps that interpolate between sparse measurements while suppressing noise \citep{carrick_2015}.
    Constrained realizations extend this framework by generating full ensemble samples of the velocity field that are statistically consistent with the observations, enabling uncertainty quantification and non-Gaussian features to be captured \citep{bertschinger_1989}.
    More recently, Bayesian approaches employing Hamiltonian Monte Carlo sampling have been developed to reconstruct the initial conditions and evolved density and velocity fields simultaneously, allowing rigorous marginalization over cosmological parameters and systematic uncertainties \citep{lilow_2021}.
    
    The conversion into a continuous 3D field makes reconstructions suitable for cosmographic mapping and characterization of individual structures such as superclusters, voids, and filaments \citep[e.g.][]{courtois_2023a}. Such detailed structural information is essential for environmental studies of galaxy formation, connecting the observed galaxy population to the underlying dark matter distribution and the cosmic web topology.
    
    Beyond cosmographic mapping, 3D velocity field reconstructions can also play an indirect role in enabling the 3D extensions of cosmological analyses presented throughout this section.
    While state-of-the-art methods such as maximum likelihood estimates often rely on direct measurements of line-of-sight peculiar velocities rather than reconstructed fields, the ability to transform discrete measurements into continuous volume-filling maps provides complementary insights.
    Furthermore, reconstructed fields facilitate cross-correlations with other cosmological probes by providing velocity estimates at arbitrary positions.

    The anticipated accessible volume and sample size of SKAO TF peculiar velocity surveys represent a transformative leap in the field.
    Moreover, the homogeneous selection function simplifies the treatment of observational biases in reconstruction algorithms, reducing systematic uncertainties that currently limit the interpretation of reconstructed fields.
    The combination of high sampling density, large volume, and uniform selection will enable SKAO to set a new standard in data-driven reconstructions of the local cosmic velocity field.

\subsection{Combination with Weak Lensing} \label{sec:WLcombinations}

    Weak gravitational lensing and peculiar velocities probe complementary aspects of the same underlying gravitational field, making their combination a powerful tool for precision cosmology and tests of General Relativity.
    Both observables are sensitive to the metric perturbations $\Phi$ and $\Psi$, but through fundamentally different physical mechanisms.
    Weak lensing measures the integrated gravitational potential along the line of sight through the convergence field,
    \begin{equation}
        \kappa(\hat{n}) \propto \int_0^{\chi_s} d\chi \, W_\kappa(\chi, \chi_s) \, (\Phi + \Psi),
    \end{equation}
    where $W_\kappa(\chi, \chi_s)$ is a lensing kernel that depends on the source redshift distribution and $\chi$ is comoving distance.
    Meanwhile, peculiar velocities are sourced by the local gravitational acceleration, with the velocity divergence in the linear regime given by
    \begin{equation}
        \nabla \cdot \mathbf{v} \propto -H(z) f(z) \delta \propto -H(z) f(z) \nabla^2 (\Phi + \Psi),
    \end{equation}
    where $f(z)$ is the growth rate and $\delta$ is the matter overdensity.
    This distinct sensitivity to the gravitational field means that cross-correlations between convergence and velocity divergence can break degeneracies that limit individual probes.

    An SKAO TF peculiar velocity survey will enable several powerful synergies with weak lensing observations from concurrent surveys such as LSST, \textit{Euclid}, and the \textit{Nancy Grace Roman Space Telescope}, or even from a commensal SKAO radio-continuum cosmic shear survey \citep{Harrison02.2026.SKA}:
    
    First, cross-correlating TF-derived peculiar velocities with lensing convergence maps provides measurements of $f\!\sigma_8$ that are independent of galaxy bias \citep{song_2011}.
    In comparison with single probe measurements or combinations discussed in previous sections, the velocity--lensing cross-correlation information could yield even better constraints, with a signal potentially easier to detect.

    Second, these cross-correlations enable constraints on the gravitational slip parameter $\eta \equiv \Phi/\Psi$, a key discriminator between General Relativity and modified gravity theories.
    In GR, the two metric potentials are equal in the absence of anisotropic stress ($\eta = 1$), but many alternative theories predict deviations from this relation \citep[see also][]{Camera01.2026.SKA}.
    Lensing is sensitive to $\Phi + \Psi$ while peculiar velocities probe a combination weighted by the growth rate, providing complementary constraints that can isolate $\eta$ when combined with galaxy clustering \citep[e.g.][]{Harrison01.2026.SKA}.

    Third, the large overlap between SKAO \HI{} detections and optical/near-infrared surveys creates opportunities for joint systematics mitigation.
    Intrinsic alignments of galaxy shapes with the large-scale structure represent a major systematic in weak lensing analyses.
    The availability of rotation velocities and inclinations from SKAO TF measurements for a subset of lensing sources allows empirical modeling of the intrinsic shape--velocity correlation, providing tighter constraints on intrinsic alignment contamination \citep[see also][]{Harrison02.2026.SKA,Tripathi01.2026.SKA}.

\subsubsection{Kinematic Lensing}

    Kinematic lensing represents a conceptually distinct approach that directly leverages the kinematic information from rotating disk galaxies to infer their intrinsic shapes and orientations \citep{huff_2013}.
    The method is based on the premise that the observed velocity field of a disk galaxy encodes its three-dimensional orientation: the kinematic major axis defines the projected rotation axis, while the amplitude of the rotation curve (measured via the Tully-Fisher relation) constrains the disk inclination.
    By comparing this kinematically inferred orientation to the observed photometric ellipticity and position angle, one can in principle separate the gravitational shear signal from the galaxy's intrinsic shape.

    Traditional weak lensing analyses rely on the assumption that galaxy orientations are randomly distributed on the sky, so that any coherent pattern in observed ellipticities reflects gravitational shear.
    However, tidal interactions with the large-scale structure can induce intrinsic alignments (IA) that mimic or dilute the lensing signal.
    Kinematic lensing bypasses this assumption by using velocity information to determine the true three-dimensional shape of each galaxy, allowing direct measurement of both lensing shear and intrinsic alignment \citep{Huang_2024}.

    For an SKAO TF survey, kinematic lensing becomes particularly powerful thanks to the combination of high-quality rotation curves and large statistical samples.
    When paired with deep optical/near-infrared imaging from e.g. LSST and \textit{Euclid}, this creates an ideal dataset for kinematic lensing.

\subsection{Combination with CMB measurements} \label{sec:CMBcombinations}

    The cosmic microwave background provides a well-studied screen against which late-time structure formation imprints secondary anisotropies through gravitational and kinematic interactions.
    Cross-correlating SKAO TF peculiar velocities with CMB observations from \textit{Planck}, the Simons Observatory, CMB-S4, and future missions opens multiple complementary avenues for cosmology.
    These synergies leverage the fact that both peculiar velocities and various CMB secondary effects trace the evolving gravitational field and matter distribution, but with different sensitivities to cosmological parameters and systematic effects.

\subsubsection{Integrated Sachs-Wolfe Effect}

    The integrated Sachs-Wolfe (ISW) effect produces large-angle CMB temperature anisotropies through the time-evolution of gravitational potentials as CMB photons traverse evolving large-scale structure.
    In a matter-dominated universe, gravitational potentials remain constant, and photons experience equal and opposite gravitational redshifts when entering and leaving potential wells, leading to no net effect.
    However, during epochs of accelerated expansion, driven by dark energy or a cosmological constant, potentials decay with time, and this decay imprints secondary temperature fluctuations in the CMB that are strongly correlated with the late-time matter distribution.
    Crucially, the ISW effect is maximally sensitive to the low-redshift Universe ($z \lesssim 1$), where SKAO TF measurements will provide high-density tracers of the large-scale structure.

    Previous ISW detections have relied on cross-correlations with galaxy density fields from photometric surveys, achieving detection significances of $\sim 3$--$4\sigma$. However, galaxy clustering measurements are plagued by galaxy bias $b_{\rm g}(z)$, which dilutes the ISW signal and introduces systematic uncertainties. This makes velocity-ISW cross-correlations a cleaner probe of late-time gravitational physics, enabling not only a secure detection of the ISW effect but also measurement of its redshift dependence, constraining the evolution of dark energy through $w(z)$.
    Moreover, the combination of ISW measurements with growth rate constraints from the same data provides consistency checks on the standard model: any tension between the inferred expansion history (from ISW) and growth history (from $f\!\sigma_8$) would signal new physics.

\subsubsection{Kinetic Sunyaev-Zeldovich Effect}

    The kinetic Sunyaev-Zeldovich (kSZ) effect arises when CMB photons scatter off electrons in ionized gas with a non-zero bulk velocity relative to the CMB rest frame.
    This Doppler shift produces a temperature decrement or increment in the CMB proportional to the line-of-sight peculiar velocity of the scattering medium and the optical depth through the gas.
    Unlike the thermal SZ effect, which depends on the electron temperature and produces a characteristic spectral distortion, the kSZ effect is spectrally indistinguishable from the primary CMB, making it challenging to isolate.

    The kSZ effect has been detected statistically through cross-correlations with large-scale structure tracers, and individual detections in massive galaxy clusters are beginning to emerge.
    However, the small amplitude of the effect ($\Delta T / T \sim 10^{-5}$ for typical cluster velocities) and confusion with primary CMB anisotropies and foregrounds have limited its utility as a cosmological probe.
    Cross-correlating CMB maps with independent peculiar velocity measurements offers a path to overcome these limitations.
    By stacking CMB temperature measurements at the positions of galaxies with known peculiar velocities, one can enhance the kSZ signal and suppress uncorrelated noise from the primary CMB and instrumental effects.
    Cross-correlation between velocities and CMB temperature fluctuations at the galaxy positions will enable several science applications:
    
    First, the kSZ-velocity cross-correlation provides yet another independent constraint on the growth rate $f\!\sigma_8$: the kSZ amplitude depends on both the velocity field and the free electron density (related to the baryon distribution), and consistency between kSZ-derived and dynamically-derived velocities validates the cosmological model.
    
    Second, the kSZ effect is sensitive to the ionization state of the intergalactic and circumgalactic medium, and strong kSZ signals in low-mass systems can probe the distribution of ionized gas around galaxies.
    By comparing the kSZ amplitude measured for different galaxy masses and environments with the velocities inferred from TF, one can map the baryon distribution and constrain feedback processes that redistribute gas within and beyond dark matter halos.
    
    Third, the combination of kSZ measurements with TF velocities allows calibration of the so-called ``kSZ-velocity'' relation, which can then be applied to the full photometric galaxy sample (without velocity measurements) to infer a statistical velocity field over much larger volumes.
    This approach converts the CMB into a velocity probe, extending the reach of peculiar velocity cosmology beyond the direct TF sample.

\section{Discussion} \label{sec:discussion}

    The SKA Observatory will provide unprecedented capabilities for Tully-Fisher \HI{} galaxy surveys.
    However, several challenges must be carefully addressed to realize the full scientific value of these measurements.
    
    Accurate calibration of the TF zero point remains the dominant systematic uncertainty for $H_0$ measurements.
    While statistical errors will shrink dramatically with sample size, calibration uncertainties will not automatically improve.
    
    This necessitates a parallel program of high-precision distance measurements to nearby calibrators using Cepheids \citep{riess_2016,riess_2022}, the tip of the red giant branch \citep[TRGB;][]{freedman_2019,freedman_2025}, or other primary indicators.
    Careful attention to consistency between local calibrators and the SKAO sample is essential.
    The recent development of an ``$H_0$ tension'' between Cepheid-based and TRGB-based calibrations underscores the importance of understanding systematic differences between primary distance indicators, and of the ongoing efforts to homogenize TF calibration across different distance anchors \citep[e.g.][]{kourkchi_2022,tully_2023}.

    Inclination corrections represent another critical systematic.
    While spatially resolved \HI{} imaging enables derivation of kinematic inclinations, potential misalignments between the optical and \HI{} disks, beam smearing effects, and uncertainties in deprojecting non-circular motions introduce scatter and potential biases.
    The degeneracy between inclination and intrinsic velocity dispersion for nearly face-on systems necessitates careful sample cuts, but such cuts must be accounted for in the selection function to avoid biasing cosmological inferences.
    Deep optical imaging from LSST and high-resolution NIR imaging from Euclid will provide independent inclination estimates, enabling cross-validation and identification of problematic systems.

    Systematic errors in velocity width measurements, arising from spectral resolution, signal-to-noise limitations, and contamination from nearby sources or RFI, must be carefully characterized.
    In this work, we try to take conservative selection function assumptions that should mitigate spurious detections, but edge effects, baseline subtraction artifacts, and asymmetric profiles can still bias velocity width estimates.
    Automated and semi-automated pipelines \citep[e.g.][]{ball_2023} must be rigorously validated on mock observations with realistic noise and RFI characteristics, with particular attention to the transition from resolved to marginally resolved line profiles at higher redshifts.
    
    While SKA-Mid's sensitivity will push to lower \HI{} masses and surface brightnesses than ever before, survey completeness and selection functions need to be properly characterized.
    To this aim, dedicated mock \HI{} galaxy catalogs, ideally multiple and based on independent models \citep[see also][]{Ronconi01.2026.SKA}, are crucial.
    The framework of \citet{mayor_2026}, building on semi-analytic models such as GAEA \citep{de_lucia_2024}, validated against observational data, and employed in our forecasts, provides a ready tool for such purpose.

    Environmental effects, i.e.\ variations of the TFr with local galaxy density, cluster membership, or cosmic web environment, could introduce systematic errors if not properly accounted for.
    While the TFr is remarkably universal across diverse environments in the local Universe, potential environmental dependencies at higher redshift or in extreme overdensities merit investigation.
    The large SKAO sample will enable explicit tests for environmental trends by comparing TFr parameters in voids, filaments, and clusters identified from the galaxy distribution or from ancillary large-scale structure tracers.

    Photometric systematics in the optical/NIR data used for luminosities and stellar masses also propagate into TF analyses.
    Dust corrections, aperture matching between radio and optical data, star formation history assumptions for stellar mass estimates, and photometric zero-point calibrations all contribute uncertainties.
    The availability of multi-band photometry from LSST and Euclid will enable robust SED fitting and dust corrections, but consistency checks between different photometric systems and cross-validation with spectroscopic stellar masses are prudent.

\section{Conclusion}

    The Tully-Fisher relation has served as a cornerstone of extragalactic astronomy for nearly five decades, evolving from a redshift-independent distance indicator into a probe of galaxy formation physics and a tracer of large-scale structure.
    The SKAO will complete this transformation, elevating the TFr into a premier tool for precision cosmology.

    The breadth of enabled science cases discussed in this chapter illustrates the richness of cosmological information encoded in galaxy peculiar velocities.
    The systematic challenges, while substantial, are tractable through careful survey design, robust calibration programs, and detailed analysis methodologies.
    Synergies between different cosmological probes and between SKAO and complementary facilities not only mitigate systematics but also amplify the scientific return, enabling combined tests of fundamental physics across wavelengths.

    The combination of unprecedented sample sizes, extended redshift reach, homogeneous data quality, and synergistic capabilities positions SKAO TF \HI{} observations as uniquely powerful measurements to improve our understanding of the Universe's large-scale structure, dynamics, and fundamental physical laws.
    The coming decades promise an extraordinarily exciting period for peculiar velocity cosmology, with the SKAO playing a central enabling role.
    
\section*{Acknowledgments}

    This work was supported in part by grant CRSII5\_193826 from the Swiss National Science Foundation, and by the Swiss State Secretariat for Education, Research and Innovation (SERI) as part of the SKACH consortium. AN and PB acknowledge support from the European Research Council (ERC) under the European Union's Horizon 2020 research and innovation programme (Grant agreement No. 948764).

\section*{Author List Ordering}

J. Mayor led the writing of this chapter, with key contributions from A. Carr and K. Said.
The conception and development of the chapter was the result discussions within the \HI{} Galaxy Focus Group of the SKA Cosmology SWG, co-led by G. De Lucia and A. Ponomareva, who also contributed to its writing and revision. The remaining authors contributed to the writing and/or the revision of this manuscript and are listed in alphabetical order.


\newpage
\bibliographystyle{abbrvnat-maxbibnames4}
\bibliography{references,references_cross_aaskaii}

\end{document}